\documentclass[a4paper,12pt]{article}
\usepackage{amsmath}
\usepackage{amssymb}
\usepackage{latexsym}
\newcommand{\be}{\begin{eqnarray}}
\newcommand{\ee}{\end{eqnarray}}
\newcommand{\ph}{\phantom}
\begin{document}
\title{\large\textbf{Automorphisms and a Cartography of the Solution Space for Vacuum
Bianchi Cosmologies: The Type III Case}}
\author{\textbf{T. Christodoulakis}\thanks{tchris@phys.uoa.gr} ~\textbf{and}~\textbf{Petros A. Terzis}\thanks{pterzis@phys.uoa.gr}\\ University of Athens, Physics Department\\
Nuclear \& Particle Physics Section\\
Panepistimioupolis, Ilisia GR 157--71, Athens, Hellas}
\date{}
\maketitle
\begin{center}
\textit{}
\end{center}
\vspace{0.5cm} \numberwithin{equation}{section}
\begin{abstract}
The theory of symmetries of systems of coupled, ordinary
differential equations (ODE's) is used to develop a concise
algorithm for cartographing the space of solutions to vacuum Bianchi
Einstein's Field Equations (EFE). The symmetries used are the well
known automorphisms of the Lie algebra for the corresponding
isometry group of each Bianchi Type, as well as the scaling and the
time reparameterization symmetry. Application of the method to Type
III results in: a) the recovery of all known solutions without prior
assumption of any extra symmetry, b) the enclosure of the entire
unknown part of the solution space into a single, second order ODE
in terms of one dependent variable and c) a partial solution to this
ODE. It is also worth-mentioning the fact that the solution space is
seen to be naturally partitioned into three distinct, disconnected
pieces: one consisting of the known Siklos (pp-wave) solution,
another occupied by the Type III member of the known Ellis-MacCallum
family and the third described by the aforementioned ODE in which an
one parameter subfamily of the known Kinnersley geometries resides.
Lastly, preliminary results reported show that the unknown part of
the solution space for other Bianchi Types is described by a
strikingly similar ODE, pointing to a natural operational
unification as far as the problem of solving the cosmological EFE's
is concerned.
\end{abstract}
\newpage
\section{Introduction}
Since the early times of cosmology, Automorphisms have  been
identified as possible key elements for a unified treatment of
spatially homogeneous Bianchi Geometries \cite{Schuk}. Harvey has
found the automorphisms of all 3-dimensional Lie Algebras
\cite{Harvey}, while the corresponding results for the 4-dimensional
Lie Algebras have been reported in \cite{ChrDim}. Jantzen's tangent
space approach sees the automorphic matrices as the means for
achieving a convenient parametrization of a full scale-factor matrix
in terms of a, desired, diagonal matrix \cite{jantzen}. Samuel and
Ashtekar were the first to look upon Automorphisms from a space
viewpoint \cite{ashtekar}. The notion of \emph{Time-Dependent
Automorphism Inducing Diffeomorphisms} (A.I.D.'s), i.e. coordinate
transformations mixing space and time in the new spatial coordinates
and inducing automorphic motions on the scale-factor matrix, the
lapse and the shift has been developed in \cite{JMP}. \\In this
communication we revisit the problem of solving the EFE's for vacuum
Bianchi Geometries. We begin with a full metric, i.e. we make no
assumption for the lapse function $N^2$, the shift vector $N^\alpha$
and the spatial metric $\gamma_{\alpha\beta}$. Then we use the
Time-Dependent A.I.D.'s to put the shift vector to zero. At this
point the idea is to exploit, in a systematic way, the remaining
symmetries of the field equations --sometimes called ''rigid''
\cite{Henneaux}-- to transform them to the most simple form
possible, without loss of generality. These are the well known
symmetries following from the constant Automorphism group within
each Bianchi Type, as well as the scaling of the metric by a
constant and the time reparameterization symmetry (see e.g.
\cite{MacBook}). Applying this analysis to Bianchi Type III Vacuum
Cosmology we produce an exhaustive cartography of the entire space
of its solutions.

The paper is organized as follows: in section 2, we present our
method.  In section 3, after a brief description of Bianchi Type III
Cosmology we apply the method. We thus recover all known solutions,
describe the unknown part of the solution space with a single,
second order ODE in terms of one dependent variable and present a
new solution. A brief preview of the corresponding results obtained
for other Bianchi Types is also included.  Finally some discussion
and concluding remarks are given in section 4.

\section{The Method}
As it is well known, for spatially homogeneous spacetimes with a
simply transitive action of the corresponding isometry group
\cite{EM}, \cite{MacBook}, the line element, assumes the form
\begin{equation}\label{line element}
ds^2=\left(N^\alpha N_\alpha-N^2\right)dt^2+2N_\alpha
\sigma^\alpha_idx^idt+\gamma_{\alpha\beta}\sigma^\alpha_i\sigma^\beta_jdx^idx^j
\end{equation}
where the 1-forms $\sigma^{\alpha}_{i}$, are defined from:
\begin{equation}\label{ορισμός σ}
d\sigma^{\alpha}=C^\alpha_{\beta\gamma}\sigma^\beta\wedge\sigma^\gamma
\Leftrightarrow \sigma^{\alpha}_{i,j} -
\sigma^\alpha_{j,i}=2C^\alpha_{\beta\gamma} \sigma^{\gamma}_{i}
\sigma^\beta_{j}.
\end{equation}
Then the field equations are (e.g. \cite{JMP}):
\begin{equation}\label{τετραγωνικός σύνδεσμος}
E_o\doteq K^{\alpha\beta}K_{\alpha\beta}-K^2- \mathbf{R}=0
\end{equation}
\begin{equation}\label{γραμμικός σύνδεσμος}
E_\alpha\doteq K^\mu_\alpha
C^\epsilon_{\mu\epsilon}-K^\mu_\epsilon C^\epsilon_{\alpha\mu}=0
\end{equation}
\begin{equation}\label{δυναμικές εξισώσεις}
E_{\alpha\beta}\doteq \dot{K}_{\alpha\beta}+N\left(2K^\tau_\alpha
K_{\tau\beta}-K K_{\alpha\beta}\right)+2N^\rho\left(K_{\alpha\nu}
C^\nu_{\beta\rho}+K_{\beta\nu} C^\nu_{\alpha\rho} \right)-N
\mathbf{R}_{\alpha\beta}=0
\end{equation}
where
\begin{equation}\label{K}
K_{\alpha\beta}=-\frac{1}{2N}\left(\dot{\gamma}_{\alpha\beta}+2\gamma_{\alpha\nu}
C^\nu_{\beta\rho} N^\rho+2\gamma_{\beta\nu} C^\nu_{\alpha\rho}
N^\rho \right)
\end{equation}
is the extrinsic curvature and
\begin{equation}\label{Rab}
\begin{array}{cc}
\mathbf{R}_{\alpha\beta}=&C^\kappa_{\sigma\tau} C^\lambda_{\mu\nu}
\gamma_{\alpha\kappa}\gamma_{\beta\lambda}\gamma^{\sigma\nu}\gamma^{\tau\mu}+
2C^\kappa_{\beta\lambda} C^\lambda_{\alpha\kappa}+ 2
C^\mu_{\alpha\kappa}
C^\nu_{\beta\lambda}\gamma_{\mu\nu}\gamma^{\kappa\lambda}+\\
&\\
 &2C^\lambda_{\beta\kappa}
C^\mu_{\mu\nu}\gamma_{\alpha\lambda}\gamma^{\kappa\nu}+ 2
C^\lambda_{\alpha\kappa}
C^\mu_{\mu\nu}\gamma_{\beta\lambda}\gamma^{\kappa\nu}
\end{array}
\end{equation}
the Ricci tensor of the hyper-surface.

In \cite{JMP} particular spacetime coordinate transformations have
been found, which reveal  as symmetries of (\ref{τετραγωνικός
σύνδεσμος}), (\ref{γραμμικός σύνδεσμος}), (\ref{δυναμικές
εξισώσεις})  the following transformations of the dependent
variables $N, N_\alpha, \gamma_{\alpha\beta}$ :
\begin{equation}\label{gaugetrans}
\tilde{N}=N, \,
\tilde{N}_{\alpha}=\Lambda^\rho_\alpha\,(N_\rho+\gamma_{\rho\sigma}\,P^\sigma),
\, \tilde{\gamma}_{\mu\nu}=\Lambda^\alpha_\mu \, \Lambda^\beta_\nu
\, \gamma_{\alpha\beta}
\end{equation}
where the matrix $\Lambda$ and the triplet $P^\alpha$ must
satisfy: \be\label{intcon}
 \Lambda^\alpha_\rho \, C^\rho_{\beta\gamma}& =
&C^\alpha_{\mu\nu}\ \Lambda^\mu_\beta \,\Lambda^\nu_\gamma\ee
\be\label{intcon1}\ 2\,P^\mu\,C^\alpha_{\mu\nu}\Lambda^\nu_\beta &=&
\dot{\Lambda}^\alpha_\beta \ee

For all Bianchi Types, this system of equations admits solutions
which contain three arbitrary functions of time plus several
constants depending on the Automorphism group of each type. The
three functions of time, are distributed among $\Lambda$ and $P$
(which also contains derivatives of these functions). So one can
use this freedom either to simplify the form of the scale factor
matrix or to set the shift vector to zero. The second action can
always be taken, since, for every Bianchi type, all three
functions appear in $P^\alpha$.

In this work we adopt the latter point of view. When the shift has
been set to zero, there is still a remaining "gauge" freedom
consisting of all constant $\Lambda^\alpha_\beta$ (Automorphism
group matrices). Indeed the system (\ref{intcon}), (\ref{intcon1})
accepts the solution $\Lambda^\alpha_\beta=constant$,
$P^\alpha=\mathbf{0}$. The generators of the corresponding motions,
induced in the space of dependent variables spanned by
$\gamma_{\alpha\beta}\,'s$ (the lapse is given in terms of
$\gamma_{\alpha\beta},\,\dot{\gamma}_{\alpha\beta}$ by algebraically
solving the quadratic constraint equation )
$\tilde{\gamma}_{\mu\nu}=\Lambda^\alpha_\mu \, \Lambda^\beta_\nu \,
\gamma_{\alpha\beta}$ are  \cite{CMP} :
\begin{equation}\label{genX}
X_{(I)}=\lambda^\rho_{(I)\alpha}\,\gamma_{\rho\beta}\,
\frac{\partial}{\partial\gamma_{\alpha\beta}}
\end{equation}
with $\lambda$ satisfying:
\begin{equation}\label{deflamda}
\lambda^\alpha_{(I)\rho}\,C^\rho_{\beta\gamma}=\lambda^\rho_{(I)\beta}\,C^\alpha_{\rho\gamma}+
\lambda^\rho_{(I)\gamma}\,C^\alpha_{\beta\rho}.
\end{equation}

Now, these generators define a Lie algebra and each one of them
induces, through its integral curves, a transformation on the
configuration space spanned by the $\gamma_{\alpha\beta}$'s. If a
generator is brought to its normal form (e.g.
$\frac{\partial}{\partial z_{i}}$), then the Einstein equations,
written in terms of the new dependent variables, will not explicitly
involve  $z_{i}$. They thus become a \emph{first order} system in
the function $\dot{z}_{i}$ \cite{Stephani}. If the above Lie algebra
happens to be abelian, then all generators can be brought, to their
normal form simultaneously. If this is not the case, we can
diagonalize in one step the generators corresponding to any eventual
abelian subgroup . The rest of the generators (not brought in their
normal form) continue to define a symmetry of the reduced system of
EFE's if the algebra of the $X_{(I)}$'s is solvable \cite{Olver}.
One can thus repeat the previous step, by choosing one of these
remaining generators. This choice will of course depend upon the
simplifications brought to the system at the previous level. Finally
if the algebra does not contain any abelian subgroup, one can always
choose one of the generators, bring it to its normal form, reduce
the system and search for its symmetries (if there are any). Lastly,
two further symmetries of (\ref{τετραγωνικός σύνδεσμος}),
(\ref{γραμμικός σύνδεσμος}), (\ref{δυναμικές εξισώσεις}) are also
present and can be used in conjunction with the constant
automorphisms: The time reparameterization $t \rightarrow
f(t)+\alpha$, owing to the non-explicit appearance of time in these
equations, and the scaling by a constant $\gamma_{\alpha\beta}
\rightarrow \mu \gamma_{\alpha\beta}$ as can be straightforwardly
verified. Their corresponding generators are: \\\be\label{ορισμός
Y1} Y_1  =\frac{1}{\dot{f}}\, \frac{\partial}{\partial t} \ee
\begin{equation}\label{ορισμός Y2}
Y_2=\gamma_{\alpha\beta}\,
\frac{\partial}{\partial\gamma_{\alpha\beta}}
\end{equation}

These generators commute among themselves, as well as with the
$X_{(I)}$'s, as it can be easily checked.

\section{Application to Bianchi Type III}
We are now going to apply the Method, previously discussed, to the
case of Bianchi Type III. For this type the structures constants
are \cite{Ryan}
\begin{equation}\label{σταθερές δομής}
\begin{array}{ll}
C^1_{13}=-C^1_{31}=1\\
C^\alpha_{\beta\gamma}=0 & for\, all\, other\, values\, of\,
\alpha \beta \gamma
\end{array}
\end{equation}
Using these values in the defining relation (\ref{ορισμός σ}) of
the 1-forms $\sigma^{\alpha}_{i}$ we obtain
\begin{equation}
\sigma^\alpha_i=\left(\begin{matrix} 0 & e^{-x} & 0 \cr 0 & 0 & 1
\cr \frac{1}{2} & 0 & 0
\end{matrix}
\right)
\end{equation}
The corresponding vector fields $\xi^i_\alpha$ ( satisfying
$[\xi_\alpha,\xi_\beta]=c^\gamma_{\alpha\beta} \xi_\gamma$ ) with
respect to which the Lie Derivative of the above 1-forms is zero
are: \be\label{killing}
\begin{array}{lll}
\xi_1=\partial_y & \xi_2=\partial_z & \xi_3=\partial_x+y\partial_y
\end{array}
\ee

The Time Depended A.I.D.'s are described by
\begin{equation}\label{Aut Λ}
 \Lambda^\alpha_\beta=
\left(\begin{matrix} e^{-2P(t)}&0&x(t)\cr 0&c_{22}&c_{23}\cr
0&0&1\end{matrix}
 \right)
\end{equation} και
\begin{equation}\label{Aut P}
P^\alpha=\left(x(t)
\dot{P}(t)+\frac{1}{2}\dot{x}(t),P^2(t),\dot{P}(t)\right)
\end{equation}
where $P(t), x(t)$ and $P^2(t)$ are arbitrary functions of time.
As  we have already remarked the three arbitrary functions appear
in $P^\alpha$ and thus can be used to set the shift vector to
zero.

The remaining symmetry of the EFE's is, consequently, described by
the constant matrix:
\begin{equation}\label{Outer Aut}
M=\left(\begin{matrix}e^{s_{1}}&0&s_{4} \cr 0&e^{s_{2}}&s_{3} \cr
0&0&1\end{matrix}\right)
\end{equation}
where the parametrization has been chosen so that the matrix becomes
identity for the zero value of all parameters.

 Thus the induced transformation on the scale factor matrix is
$\tilde{\gamma}_{\alpha\beta}=M^{\mu}_{\alpha}M^{\nu}_{\beta}\gamma_{\mu\nu}$,
which explicitly reads:
\begin{equation}\label{gamma new}
\left\{
\begin{array}{l}
\tilde{\gamma}_{11}=e^{2\,{s_1}}\,{{\gamma }_{11}}\\
\\
\tilde{\gamma}_{12}=e^{{s_1} + {s_2}}\,{{\gamma }_{12}}\\
\\
\tilde{\gamma}_{13}=e^{{s_1}}\,\left( {s_3}\,{{\gamma }_{11}} + {s_4}\,{{\gamma }_{12}} + {{\gamma }_{13}} \right)\\
\\
\tilde{\gamma}_{22}=e^{2\,{s_2}}\,{{\gamma }_{22}}\\
\\
\tilde{\gamma}_{23}=e^{{s_2}}\,\left( {s_3}\,{{\gamma }_{12}} +
{s_4}\,{{\gamma }_{22}} + {{\gamma }_{23}} \right)\\
\\
\tilde{\gamma}_{33}={{s_3}}^2\,{{\gamma }_{11}} + 2\,{s_3}\,
   \left( {s_4}\,{{\gamma }_{12}} + {{\gamma }_{13}} \right)  + {{s_4}}^2\,{{\gamma }_{22}} +
  2\,{s_4}\,{{\gamma }_{23}} + {{\gamma }_{33}}
\end{array}\right.
\end{equation}

The previous equations, define a group of transformations $G_{r}$
of dimension  $r=dim(Aut(III))=4$. The four generators of the
group, can be evaluated from the relation:
\begin{equation}\label{ορισμός Χ}
X_{A}=\left(\frac{\partial\tilde{\gamma}_{\alpha\beta}}{\partial
s_{A}}\right)_{s=0}\frac{\partial}{\partial\gamma_{\alpha\beta}}
\end{equation}
where $A=\left\{1,2,3,4\right\}$. Applying this definition to
(\ref{gamma new}) we have the generators:
\begin{equation}\label{X1}
X_{1}=2\gamma_{11}\frac{\partial}{\partial\gamma_{11}}+\gamma_{12}\frac{\partial}{\partial\gamma_{12}}
+\gamma_{13}\frac{\partial}{\partial\gamma_{13}}
\end{equation}
\begin{equation}\label{X2}
X_{2}=\gamma_{12}\frac{\partial}{\partial\gamma_{12}}+2\gamma_{22}\frac{\partial}{\partial\gamma_{22}}
+\gamma_{23}\frac{\partial}{\partial\gamma_{23}}
\end{equation}
\begin{equation}\label{X3}
X_{3}=\gamma_{12}\frac{\partial}{\partial\gamma_{13}}+\gamma_{22}\frac{\partial}{\partial\gamma_{23}}
+2\gamma_{23}\frac{\partial}{\partial\gamma_{33}}
\end{equation}
\begin{equation}\label{X4}
X_{4}=\gamma_{11}\frac{\partial}{\partial\gamma_{13}}+\gamma_{12}\frac{\partial}{\partial\gamma_{23}}
+2\gamma_{13}\frac{\partial}{\partial\gamma_{33}}
\end{equation}

The algebra $\textsl{g}_{r}$ that corresponds to the group $G_{r}$
has the following table of commutators:
\begin{equation}\label{μεταθέτες}
\begin{array}{lll}
\left[X_{1},X_{2}\right]=0,&\left[X_{1},X_{3}\right]=0,&\left[X_{1},X_{4}\right]=X_{4},
\\ \left[X_{2},X_{3}\right]=X_{3},& \left[X_{2},X_{4}\right]=0,&\left[X_{3},X_{4}\right]=0
\end{array}
\end{equation}

As it is evident from the above commutators (\ref{μεταθέτες})  the
group is non-abelian, so we cannot diagonalize  at the same time all
the generators. However, if we calculate the derived algebra of
$\textsl{g}_{r}$, we have
\begin{equation}
\textsl{g}_{r'}=\left\{[X_{A},X_{B}]: X_{A}, X_{B}\in
\textsl{g}_{r}\right\}\Rightarrow
\textsl{g}_{r'}=\left\{X_{3},X_{4}\right\}
\end{equation}
and furthermore, it's second derived algebra reads:
\begin{equation}
\textsl{g}_{r''}=\left\{[X_{A},X_{B}]: X_{A}, X_{B}\in
\textsl{g}_{r'}\right\}\Rightarrow
\textsl{g}_{r''}=\left\{0\right\}
\end{equation}

Thus, the group  $G_{r}$ is solvable since the $\textsl{g}_{r''}$ is
zero. As it is evident  $X_{3}, X_{4}, Y_{2}$ generate an Abelian
subgroup, and we can, therefore,  bring them to their normal form
simultaneously. The appropriate transformation of the dependent
variables is: \be\label{gammav} \left\{
\begin{array}{l}
\gamma_{11}= e^{{u_1} + 2\,{u_6}} \\
\\
\gamma_{12}=e^{{u_1} + {u_2} + {u_4} + {u_6}}\\
\\
\gamma_{13}=e^{{u_1} + {u_6}}\,\left( e^{{u_6}}\,{u_3} + e^{{u_2}
+ {u_4}}\,{u_5} \right) \\
\\
\gamma_{22}=e^{{u_1} +
2\,{u_4}}\\
\\
\gamma_{23}= e^{{u_1} + {u_4}}\,\left( e^{{u_2} + {u_6}}\,{u_3} +
e^{{u_4}}\,{u_5} \right) \\
\\
\gamma_{33}=e^{{u_1}}\,\left( 1 + e^{2\,{u_6}}\,{{u_3}}^2 +
2\,e^{{u_2} + {u_4} + {u_6}}\,{u_3}\,{u_5} +
    e^{2\,{u_4}}\,{{u_5}}^2 \right)
\end{array}\right.
\ee

In these coordinates the generators $Y_{2}, X_{A}$  assume the form:

\be\label{generatorsv}
\begin{array}{lll}
Y_2=\frac{\partial}{\partial u_1} & X_3=\frac{\partial}{\partial
u_3} & X_4=\frac{\partial}{\partial u_5}  \\
\\
X_2=\frac{\partial}{\partial u_4} -u_5 \frac{\partial}{\partial
u_5}& X_1 =\frac{\partial}{\partial u_6}-u_3
\frac{\partial}{\partial u_3}
\end{array}
\ee \\Except of the parametrization ( \ref{gammav}) there is also
another one achieving the same result (\ref{generatorsv}), which
simply attributes a - sign to $\gamma_{12}$ and therefore any
solution later described will remain valid under this change.
\\Evidently, a first look at (\ref{gammav}) gives the feeling that
it would be hopeless even to write down the Einstein equation.
However, the simple form of the generators (\ref{generatorsv})
ensures us that these equations will be of first order in the
functions $\dot{u}_1$,  $\dot{u}_3$ and $\dot{u}_5$.

\subsection{Description of the Solution Space}

Before we begin solving the Einstein equations, a few comments for
the possible values of the functions $u_{i}, i={1,\ldots,6}$ will
prove very useful.

The determinant of  $\gamma_{\alpha\beta}$, is
\begin{equation}\label{detgamma}
det[\gamma_{\alpha\beta}]=e^{3\,{u_1} + 2\,\left( {u_4} + {u_6}
\right) }\,\left( 1 - e^{2\,{u_2}} \right)
\end{equation}
so we must have $u_{2}< 0$ .

The transformation from the $\gamma$' s to the $u$' s, becomes
singular when $\gamma_{12}\,=\,0$, since the function $u_2$ equals
to
\begin{equation}
u_2=\ln (|{{\gamma }_{12}}|) - \frac{\ln ({{\gamma }_{11}}\,{{\gamma
}_{22}})}{2}.
\end{equation}
So two cases are naturally arising, according to whether
$\gamma_{12}$ is different or equal to zero. \\If
$\gamma_{12}\,\neq\,0$ the two linear constraint equations, written
in the new variables (\ref{gammav}), give \be E_1=0 \Rightarrow
- e^{{u_6}}\,\left( e^{{u_6}}\,{\dot{u}_3} + e^{{u_2} + {u_4}}\,{\dot{u}_5} \right)  =0 \\
E_2=0\Rightarrow -\frac{1}{2}\,e^{{u_4}}\,\left( e^{{u_2} +
{u_6}}\,{\dot{u}_3} + e^{{u_4}}\,{\dot{u}_5} \right)=0 \ee This
system admits only the trivial solution, since the determinant of
the 2x2 matrix formed by the coefficients of $\dot{u}_3, \dot{u}_5$
becomes zero only for the forbidden value $u_2=0$. We thus have \be
u_3=k_3, & u_5=k_5 \ee Now, these values of $u_3,u_5$ make
$\gamma_{13},\gamma_{23}$ functionally dependent upon
$\gamma_{11},\gamma_{12},\gamma_{22}$ (see (\ref{gammav})). It is
thus possible to set these two components to zero by means of an
appropriate constant automorphism.\\In the case $\gamma_{12}\,=\,0$
we can again bring simultaneously into normal form the corresponding
$X_3,X_4,Y_2$ . The appropriate change of dependent variables is
given by:\be \gamma_{\alpha\beta}=\left(
\begin{matrix}
e^{u_1+2\,u_6} & 0 & e^{u_1+2\,u_6}\,u_3 \cr 0 & e^{u_1+2\,u_5} &
e^{u_1-u_4+u_5} \cr e^{u_1+2\,u_6}\,u_3 & e^{u_1-u_4+u_5} &
e^{u_1}\,(1+e^{-2\,u_4}+e^{2\,u_6}\,u_3^2)
\end{matrix} \right)
\ee In these variables all three linear constraint equations can be
integrated, yielding: \be E_1=0\Rightarrow
-e^{2\,u_6}\,\dot{u}_3=0\Rightarrow u_3=k_3 \ee \be E_2=0\Rightarrow
-\frac{1}{2}\,e^{-u_4+u_5}\,(\dot{u}_4+\dot{u}_5)=0\Rightarrow
u_5=k_5-u_4 \ee \be
E_3=0\Rightarrow\,-2\,e^{2\,u_4+2\,u_6}\,u_3\dot{u}_3+\dot{u}_4+\dot{u}_5+2\,
e^{2\,u_4}\,\dot{u}_6=0 \Rightarrow u_6=k_6 \ee. Again, these values
imply that a constant automorphism suffices to set the (13) and (23)
components of the scale-factor matrix to zero, i.e. to put it into
diagonal form. We have thus reached a first important conclusion,
that is:

\emph{Without loss of generality, we can start our investigation of
the solution space for Type III vacuum Bianchi Cosmology from a
block-diagonal form of the scale-factor matrix ( and, of course,
zero shift)} \be\label{gammau1} \gamma_{\alpha\beta}=\left(
\begin{matrix}
\gamma_{11} & \gamma_{12} & 0 \cr \gamma_{12} & \gamma_{22} & 0 \cr
0 & 0 & \gamma_{33}
\end{matrix}
\right) \ee Note that this conclusion could have not been reached
off mass-shell, due to the fact that the time-dependent Automorphism
(\ref{Aut Λ}) does not contain the necessary two arbitrary functions
of time in the (13) and (23) components ( besides the fact that all
the freedom in arbitrary functions of time has been used to set the
shift to zero). As we have earlier remarked, since the algebra
(\ref{μεταθέτες}) is solvable, the remaining (reduced) generators
$X_1,X_2$ (corresponding to diagonal constant automorphisms) as well
as $Y_2$ continue to define a Lie-Point symmetry of the reduced
EFE's and can thus be used for further integration of this system of
equations.
\subsubsection{Case I: $\gamma_{12}\,=\,0$}
The remaining (reduced) automorphism generators are
\begin{equation}
X_{1}=2\gamma_{11}\frac{\partial}{\partial\gamma_{11}},\,X_{2}=2\gamma_{22}\frac{\partial}{\partial\gamma_{22}}\nonumber
\end{equation} The appropriate change of dependent variables which brings these
generators -along with $Y_2$- into normal form, is described by the
following scale-factor matrix :\be \gamma_{\alpha\beta}=\left(
\begin{matrix}
e^{u_1+\,u_3} & 0 & 0 \cr 0 & e^{u_2+\,u_3} & 0 \cr 0 & 0 & e^{u_3}
\end{matrix} \right)
\ee In these variables the first two linear constraint equations are
identically satisfied, while the third reads $ E_3=0\Rightarrow
-2\,\dot{u}_1=0\Rightarrow u_1=k_1 $. Substituting this value of
$u_1$ into the quadratic constraint equation $E_0$ we obtain the
lapse function  \be\label{lapse1}
 N^2=\frac{1}{16}\,e^{{u_3}}\,\dot{u}_3(2\dot{u}_2+3\dot{u}_3)\ee.
 Now, substitution of $ u_1=k_1 $ and the above value for the lapse
 $N^2$ into the spatial EFE's results in the single, independent
 equation :\be\label{diag}(\dot{u}_2+\dot{u}_3)(2\dot{u}_3\ddot{u}_2
 -2\dot{u}_2\ddot{u}_3+2\dot{u}_2^2\dot{u}_3+3\dot{u}_3^2+5\dot{u}_2\dot{u}_3^2) \ee
This equation is, as expected from the theory, of the first order
in $\dot{u}_2,\dot{u}_3$. Notice that this result could have not
been reached had we chosen any particular time gauge, such as
$N^2=F(u_2,u_3,t)$ : Not only $u_2,u_3,t$ would appear in the
Spatial EFE's, but also the number of independent such equations
would have been increased to 2. This remark should not be taken as
a negative view for complete gauge fixing, but rather as pointing
to the fact that keeping the gauge freedom into the game helps
manifesting the symmetries of the system and eventually solving
the equations.\\Equation (\ref{diag}) is readily integrated,
leading to two different space-times according to which
parenthesis is set to zero. If the first is made to vanish, i.e.
$u_2=k_2-u_3$, the ensuing line-element is the known (Type III)
cosmological disguise of Minkowski space-time (\cite{Wainr}):
\be\label{metric flat} ds^2=
-\frac{1}{16}e^{{u_3}}\dot{u}_3^2\,dt^2+\frac{1}{4}e^{{u_3}}\,dx^2
+e^{k_1+u_3-2x}\,dy^2+e^{{k_2}}\, dz^2 \ee the constants being of
course absorbable by the constant automorphisms and a shift in
$u_3$.\\ If the second parenthesis of (\ref{diag}) is set to zero,
i.e. $u_2=k_3-\frac{3u_3}{2}+\ln(1+k_2\,e^{\frac{u_3}{2}})$, we
obtain an equivalent form of the Type III member of the known
Ellis-MacCallum family of solutions (\cite{MacBook},\cite{Wainr}):
\be\label{macdiag}
ds^2=\kappa^2\,\left(-\frac{e^{\frac{3u_3}{2}}\dot{u}_3^2}{4(e^{\frac{u_3}{2}}-1)}\,dt^2
+e^{{u_3}}\,dx^2+e^{u_3-2x}\,dy^2+e^{\frac{-u_3}{2}}(e^{\frac{u_3}{2}}-1)\,dz^2\right)\ee
where again we have used constant automorphisms and a shift of
$u_3$ to take outside of the metric an overall constant. We can be
assured that the constant is essential either by checking that
indeed the metric inside the parenthesis does not admit a
homothetic Killing vector field or, more primarily, by finding an
invariant relation between curvature and higher derivative
curvature scalars which explicitly involves $\kappa$. For metric
(\ref{macdiag}) one such invariant relation is:
\be\label{invar1}\frac{18\,Q_1^4}{(Q_2-\frac{g^{AB}Q_{1;A}Q_{1;B}}{Q_1})^3}=
\kappa^2,\,Q_1=R^{KLMN}R_{KLMN},\,Q_2=\Box{R^{KLMN}R_{KLMN}} \ee
where capital Latin letters denote space-time indices ranging in
the interval (0-3), the semicolon stands for covariant
differentiation, and the
 $\Box$ for the covariant D'Alebertian.
\\This relation, being a constant scalar constructed out of the
intrinsic geometry ( the Riemmann tensor and its covariant
derivatives ), characterizes , along with many others that can be
found, this metric: It will be valid for any equivalent, under
general coordinate transformations, form of (\ref{macdiag}). It is
also noteworthy to observe that in both the above line-elements the
arbitrary function of time $u_3$ appears; This is because the number
of symmetry generators matches the number of scale-factors (both are
3), so that the system of spatial EFE's is reduced to first order
without any choice of time. In the case of a block-diagonal
scale-factor matrix one of the four scale factors will have to play
the role of time before the corresponding system can be reduced.
Lastly, metric (\ref{macdiag}) admits, except of (\ref{killing}), a
fourth Killing vector field acting on the surfaces of simultaneity,
namely \be\label{mackill}
\xi_4=-2\,y\,\partial_x+(e^{2\,x}-\,y^2)\,\partial_y \ee \\There is
thus a $G_4$ symmetry group acting (of course, multiply
transitively) on each $V_3$ of this metric, with an algebra having
the following table of (non-vanishing) commutators:
\begin{equation}\label{commu1}
[\xi_1,\xi_3]=\xi_1,\,[\xi_1,\xi_4]=-2\,\xi_3,\,[\xi_3,\xi_4]=\xi_4
\end{equation}
 However, it is interesting to note that we have not
imposed the extra symmetry from the beginning , but rather it
emerged as a result of the investigation process.

\subsubsection{Case II: $\gamma_{12}\,\neq\,0$}
The remaining (reduced) automorphism generators are

\begin{equation}
X_{1}=2\gamma_{11}\frac{\partial}{\partial\gamma_{11}}+\gamma_{12}\frac{\partial}{\partial\gamma_{12}}
,\,X_{2}=\gamma_{12}\frac{\partial}{\partial\gamma_{12}}+2\gamma_{22}\frac{\partial}{\partial\gamma_{22}}\nonumber
\end{equation}
The appropriate change of dependent variables which brings these
generators -along with $Y_2$- into normal form, is now given by:
\be\label{gamnodiag} \gamma_{\alpha\beta}=\left(
\begin{matrix}
e^{{u_1} + 2\,{u_4}} & e^{{u_1} + {u_2} + {u_4}} & 0 \cr e^{{u_1} +
{u_2} + {u_4}} & e^{{u_1} + 2\,{u_2}}\,{u_3} & 0 \cr 0 & 0 &
e^{{u_1}}
\end{matrix}
\right) \ee The generators are now reduced to \be
Y_2=\frac{\partial}{\partial u_1}, \, X_2=\frac{\partial}{\partial
u_2}, \, X_1=\frac{\partial}{\partial u_4} \ee indicating that the
system will be of first order in the derivatives of these variables.
The remaining variable $u_3$ will enter, (along with
$\dot{u}_3,\,\ddot{u}_3$ ) explicitly in the system and is therefore
advisable (if not mandatory) to be used as the time parameter, i.e.
to effect the change of time coordinate \be t\rightarrow u_3(t)=s,
\, u_1(t)\rightarrow u_1(t(s)), \, u_2(t)\rightarrow u_2(t(s)), \,
u_4(t)\rightarrow u_4(t(s)). \ee This choice of time will of course
be valid only if $u_3$ is not a constant. We are thus led to
consider two cases according to the constancy or non-constancy of
this variable.
\\
\\
\noindent \textbf{The case $u_3=k_3$}\\
\\The determinant of the scale-factor matrix becomes $
det[\gamma_{\alpha\beta}]=e^{3\,{u_1} + 2\,\left( {u_2} + {u_4}
\right) }\,\left( -1 + {k_3} \right)$. We thus have $k_3>1$. The two
linear constraint equations are identically satisfied, while the
third yields  \be E_3=0 \Rightarrow \frac{\dot{u}_2+(1-2
k_3)\,\dot{u}_4}{2(1-k_3)}=0\Rightarrow
u_4=k_4+\frac{u_2}{2k_3-1}\nonumber\ee Inserting these values of
$u_3,\,u_4$ into the quadratic constraint equation we obtain the
following lapse \be\label{lapse1}
 (N)^2=\frac{e^{{u_1}}\,\left( -1 + {k_3} \right)
\,\left( 3\,{\left( 1 - 2\,{k_3} \right) }^2\,{{{{\dot{u}}}_1}}^2 +
8\,{k_3}\,\left( -1 + 2\,{k_3} \right) \,{\dot{u}_1}\,{{\dot{u}}_2}
+ 4\,{k_3}\,{{{{\dot{u}}}_2}}^2 \right) }{4\,{\left( 1 - 2\,{k_3}
\right) }^2\, \left( -3 + 4\,{k_3} \right) }\ee Use of these values
of $u_3,\,u_4,\,(N)^2$ in the spatial EFE's results, as expected, in
a system which is of the first order in the unknown variables
$\dot{u}_1,\, \dot{u}_2$. The coefficient of $\ddot{u}_2$ in
$E_{33}=0$ is \be\frac{2\,e^{{u_1}}\,{k_3}\,\dot{u}_1\,
    \left( \left( -1 + 2\,{k_3} \right) \,\dot{u}_1 + \dot{u}_2 \right) }{3\,
     {\left( 1 - 2\,{k_3} \right) }^2\,{\dot{u}_1}^2 +
    8\,{k_3}\,\left( -1 + 2\,{k_3} \right) \,
    \dot{u}_1\,\dot{u}_2 + 4\,{k_3}\, {\dot{u}_2}^2}\nonumber\ee and
can be safely regarded different from zero, since the possibilities
$\dot{u}_1=0,\,\dot{u}_2= ( 1 - 2\,k_3)\,\dot{u}_1$ easily lead
(through $E_{33}=0$ itself) to zero and negative lapse,
respectively. We can thus solve $E_{33}=0$ for $\ddot{u}_2$ and
substitute into $E_{12}=0$ which becomes  \be
 \frac{e^{k_4 + {u_1} + \frac{2\,{k_3}\,{u_2}}{-1 + 2\,{k_3}}}\,{k_3}\,
      \left( {{{\dot{u}}}_1} + 2\,{{{\dot{u}}}_2} \right) \,
      \left( \left( -3 + 6\,{k_3} \right) \,{{{\dot{u}}}_1} +
        2\,\left( 3 - 2\,{k_3} \right) \,{{{\dot{u}}}_2} \right) }{6 - 20\,{k_3} + 16\,{{k_3}}^2}
    =0 \nonumber\ee Again, the second parenthesis in the numerator
    leads to zero lapse, leaving us with the only alternative $
\dot{u}_1=-2\dot{u}_2\Rightarrow u_1=2 k_1-2u_2 $ which indeed
satisfies all spatial EFE's. Finally, inserting these values of
$u_3,\,u_4,\,u_1$ in the lapse (\ref{lapse1}) and the scale-factor
matrix (\ref{gamnodiag}) we obtain the following line-element (after
using the constant automorphisms and a shift in $u_2$ to purify the
metric from the absorbable constants) :\be\label{metric u4=con}
ds^2=-\lambda^2\,d\xi^2+\frac{\xi^2}{4}\,dx^2+e^{-2x}\xi^{4\lambda}\,dy^2
+\frac{\lambda-1}{2\lambda-1}\,dz^2+2e^{-x}\xi^{2\lambda}\,dy\,dz
\ee where the constant $\lambda$ is related to $k_3$ by $
k_3=\frac{\lambda-1}{2\lambda-1} \Rightarrow 0<\lambda<\frac{1}{2}$
and we have adopted the time gauge $e^{-u_2}=\xi$ for simplicity.
\\This metric is an equivalent form of a solution originally given
by Siklos \cite{Siklos} and reproduced in \cite{Wainr}. An overall
multiplicative constant has been omitted from (\ref{metric u4=con})
since it admits the following Homothetic Killing vector field
($\mathcal{L}_H\,g_{AB}=\mu\,g_{AB}$) \be
H^A=\xi\frac{\partial}{\partial\xi}+(1-2\lambda)y\frac{\partial}{\partial
y}+z\frac{\partial}{\partial z}\nonumber \ee It also admits three
more Killing vector fields ( except (\ref{killing})) acting on
space-time, namely \be\label{killing vectors v}  \xi_4 & = &
e^{-\frac{x}{2\,\lambda}}\,\partial_\xi
+\frac{2\,\lambda}{\xi}\,e^{-\frac{x}{2\,\lambda}}\,\partial_x
\nonumber \\
 \xi_5 & = & e^{-\frac{x}{2\,\lambda}}\,y\,\partial_\xi+
\frac{2\,\lambda\,y}{\xi}\,e^{-\frac{x}{2\,\lambda}}\,\partial_x
+\frac{\lambda(\lambda-1)}{4\lambda-1}\,e^{\frac{4\lambda-1}{2\lambda}\,x}\,\xi^{-4\lambda+1}\,\partial_y
\nonumber\\
& &
-\lambda\,e^{\frac{2\lambda-1}{2\lambda}\,x}\,\xi^{-2\lambda+1}\partial_z
\nonumber \\
\xi_6 & = &
e^{-\frac{x}{2\,\lambda}}\,z\,\partial_\xi+\frac{2\,\lambda\,z}{\xi}\,
e^{-\frac{x}{2\,\lambda}}\partial_x-\lambda\,e^{\frac{2\lambda-1}{2\lambda}\,x}\,
\xi^{-2\lambda+1}\,\partial_y \nonumber \\
& &
-\lambda\,(2\lambda-1)\,e^{-\frac{x}{2\,\lambda}}\,\xi\,\partial_z
 \nonumber\ee

The $\xi_5$ field breaks down for $\lambda=\frac{1}{4}$, in which
case the valid expression is \be \xi_5^{'}=\{
\frac{y}{e^{2\,x}},\frac{y}{2\,e^{2\,x}\,\xi },
  \frac{3\,\left( -2\,x + \log (\xi ) \right) }
   {16},\frac{-{\sqrt{\xi }}}{4\,e^x} \}\ee

 The first of these is null $\xi_4^A\,\xi_4^B\,g_{AB}=0$ and covariantly
constant $\xi^A_{4\,;B}=0$, signaling that the metric is a pp-wave.
Consequently all scalar curvatures, constructed by forming scalar
contractions of tensor product of the Riemmann tensor and its
covariant derivatives of any order (such as $Q_1,Q_2$ in
(\ref{invar1})), vanish identically (see e.g. \cite{Schmidt}). This
raises the interesting question of how can we be certain that the
constant $\lambda$ is essential. An answer can be found in terms of
equalities between tensors --constructed out of the Riemmann tensor
and its covariant derivatives-- that hold true in these space-times
\cite{Kundt},\cite{Geroch}. For metric (\ref{metric u4=con}) such a
relation is : \be\label{invar2}
R^A\,_B\,^C\,_D\,R_{AECF;G;H}=\frac{4\lambda^2-2\lambda+3}{-4\lambda^2+2\lambda+2}\,R^A\,_B\,^C\,_{D;E}\,R_{AFCG;H}\ee
By the quotient law the expression of $\lambda$ in the right-hand
side of this relation is a scalar function, and being a constant it
can not change value under any coordinate transformation; thus
$\lambda$ can not be altered by such a transformation and is,
therefore, essential.

The algebra of the six killing fields, (\ref{killing}),
(\ref{killing vectors v}) has the following table of non vanishing
commutators: \be [\xi_1,\xi_3]=\xi_1 & [\xi_1,\xi_5]=\xi_4  \quad
[\xi_3,\xi_4]=-\frac{\xi_4}{2\lambda}\quad
[\xi_3,\xi_5]=\frac{2\lambda-1}{2\lambda}\,\xi_5 &
[\xi_3,\xi_6]=\frac{-1}{2\lambda}\,\xi_6 \ee There is an isotropy
group $G_2$ of null rotations emanating from this algebra, which is
easily seen by taking a linear combination of these fields:

\be Y_1=\xi_1-2\,\lambda\,\xi_3 & Y_2=\xi_4 & Y_3=-\xi_6 \nonumber\\
Y_4=\xi_2 & Y_5=\xi_3 & Y_6=\xi_5 \ee

e.g. $[Y_1,Y_2]=Y_2$ and $[Y_1,Y_3]=Y_3$

The space (being pp-wave) does not obviously have curvature
singularities, it thus seems to be geodesically complete and is of
Petrov Type N.
\\
\\
\noindent \textbf{The case $u_3\,\neq\,k_3$}\\
\\The function $u_3$ is now a valid choice of time and  $ det[\gamma_{\alpha\beta}]=e^{3\,{u_1} + 2\,\left(
{u_2} + {u_6} \right) }\,\left( -1 + s \right) $ implies the range
$(1,+\infty)$ for the new time s. The only non-vanishing linear
constraint equation $E_3=0$ yields \be\label{u4}
u_4=\int\frac{\dot{u}_2}{2s-1}\,ds+k_4 \ee while the quadratic
constraint equation $E_0=0$ gives the lapse
\begin{equation}
\begin{split}
(N)^2 = & \frac{e^{u_1}}{4\,
    {\left( 1 - 2\,s \right) }^2\,\left( -3 + 4\,s \right)
    }\,\left[2\,(2\,s-1)^2\,\dot{u}_1+3\,(2\,s-1)^2\,(s-1)\,\dot{u}_1^2 \right.
    \\ \\
  & \left.+(4\,s-2)\,\dot{u}_2+ 8\,s\,\,(s-1)\,(2\,s-1)\,\dot{u}_1\,\dot{u}_2+
  4\,s\,(s-1)\,\dot{u}_2^2\right]
    \end{split}   \end{equation} If we insert these values
    $(N)^2\,,u_4$ into the spatial EFE's they become the following
    polynomial system of first order in $\dot{u}_1,\,\dot{u}_2$
\be\label{u1'', u2''} \ddot{u}_1=\left( 1 \, \dot{u}_1 \,
\dot{u}_1^2 \, \dot{u}_1^3  \right)\, A_1\, \left( \begin{matrix} 1
\cr \dot{u}_2 \cr  \dot{u}_2^2  \cr \dot{u}_2^3 \end{matrix}
\right)\, , \,\ddot{u}_2=\left( 1 \, \dot{u}_1 \, \dot{u}_1^2 \,
\dot{u}_1^3 \right)\, A_2\, \left( \begin{matrix} 1 \cr  \dot{u}_2
\cr \dot{u}_2^2 \cr  \dot{u}_2^3 \end{matrix} \right) \ee με
\be\label{A1} A_1=\left(\begin{matrix} 0 & \frac{2}{4\,s^2-7\,s+3} &
\frac{4\,s}{8\,s^2-10\,s+3} & 0 \cr \cr \frac{1}{4\,s^2-7\,s+3} & 4
& \frac{8\,s\,(2\,s-3)\,(s-1)}{8\,s^2+10\,s-3} & 0 \cr \cr
\frac{2\,s-3}{4\,s-3} & -\frac{16\,s^2\,(s-1)}{8\,s^2-10\,s+3} & 0 &
0 \cr \cr -\frac{6\,s\,(s-1)}{4\,s-3} & 0 & 0 & 0 \end{matrix}
\right)\ee \be\label{A2} A_2=\left(\begin{matrix} 0 &
\frac{-8\,s+5}{8\,s^3-18\,s^2+13\,s-3} &
\frac{24\,s^2-50\,s+18}{8\,s^2-10\,s+3} &
\frac{8\,s\,(2\,s-3)(s-1)}{8\,s^2-10\,s+3} \cr \cr
\frac{-4\,s+2}{4\,s^2-7\,s+3} & \frac{12\,s}{-2\,s+3} &
-\frac{16\,s^2\,(s-1)}{8\,s^2+10\,s-3} & 0 \cr \cr
\frac{-6\,s+3}{4\,s-3} & -\frac{6\,s\,(s-1)}{4\,s-3} & 0 & 0 \cr \cr
0 & 0 & 0 & 0
\end{matrix} \right)\ee
Due to the form of $A_1,\,A_2$ (their components are rational
functions of the time s), system (\ref{u1'', u2''}) can be partially
integrated with the help of the following Lie-B\"{a}klund
transformation
\begin{equation}\label{u1', u2'}
\begin{split} \dot{u}_1(s)  = &
\frac{(2\,s-3)\,\tan{r(s)}-2\,s\,(8\,s^2-10\,s+3)\,\dot{r}(s)}{4\,s\,\sqrt{s-1}\,(4\,s-3)}
\\ \\
\dot{u}_2(s) = & \frac{2\,s-1}{8\,s(4\,s-3)\,\sqrt{(s-1)^3}}\,\left(
2\,(-4\,s+3)\,\sqrt{s-1}+3\,(s-1)\,\tan{r(s)} \right.
\\ \\
&\left.+2\,s\,(s-1)\,(4\,s-3)\,\dot{r}(s)\right)
\end{split}
\end{equation}
resulting in the single, second order ODE for the variable $r(s)$
\be\label{equation r} \ddot{r} & = &
\left(\tan{r}-\frac{\sqrt{s-1}}{2}\right)\,\dot{r}^2+
\frac{(-16\,s+6)\,\sqrt{s-1}+(5\,s-3)\,\tan{r}}{2\,s\,(4\,s-3)\,\sqrt(s-1)}
\,\dot{r} \nonumber \\
&
&+\frac{-9\,(s-1)^2\,\tan^2{r}+18\,(s-1)^{3/2}\,\tan{r}+4\,s\,(4\,s-3)}{8\,s^2\,(4\,s-3)^2\,(s-1)^{3/2}}
\ee This equation contains all the information concerning the
unknown part of the solution space of the Type III vacuum
Cosmology. Unfortunately, it does not posses any Lie-point
symmetries that can be used to reduce its order and ultimately
solve it. However, its form can be substantially simplified
through the use of new dependent and independent variable
$(\rho,u(\rho))$ according to  $
r(s)=\pm\arcsin{\frac{u(\rho)}{\sqrt{\rho^2-1}}},\,
s=\frac{3\,(\rho-1)}{3\,\rho-5},\, \rho>\frac{5}{3} $ thereby
obtaining the equation \be\label{final u III}
  \ddot{u}=\pm\frac{1-\dot{u}^2}{\sqrt{(6\,\rho-10)\,(\rho^2-u^2-1)}}
  \Rightarrow \ddot{u}^2=\frac{(1-\dot{u}^2)^2}{(6\,
  \rho-10)\,(\rho^2-u^2-1)}
  \ee with the corresponding lapse \be
(N)^2=\frac{\dot{u}^2-1}{8\,(3\,\rho-5)\,(\rho^2-u^2-1)}\,e^{u_1}
\ee ($\dot{u}=\frac{du}{d\rho}$) and the scale-factor matrix is
given by (\ref{gamnodiag}) after insertion of (\ref{u4}),
$u_3=s=\frac{3\,(\rho-1)}{3\,\rho-5}$ and the transformations of
$u_1,\,u_2$ that led to $u$. Independently of the way we have
reached this result, one can check (through an algebraic computing
facility such as Mathematica) that the line element thus described
is indeed a solution of all the EFE's, provided of course
(\ref{final u III}) is satisfied. One can also check that it does
not admit any Homothetic or null, covariantly constant vector
field. Therefore, the two independent constants of the general
solution to (\ref{final u III}) along with a multiplicative
constant will comprise the expected three essential constants of
the general Type III vacuum Cosmology: The general algorithm for
calculating this number when a space time gauge has been chosen
(usually  zero-shift and unit lapse), in which case the
constraints must be viewed as restrictions on the initial data,
reads as \cite{Wainr} :
\\12 ( for the six components of $\gamma_{\alpha\beta}$) -1 ( for
the time reparameterization covariance )- number of independent
constraints -dimension of Automorphism Group.\\When a space-time
gauge has not been fixed, i.e. when constraints are being viewed
as symmetry generators, the relevant counting is given by
\cite{ChrHerv}:
\begin{equation}
\begin{split}
D & =\,  2\times \left(\text{number of}\, \gamma_{\alpha\beta} \right)\\
& -2 \times \left( \text{number of linear constraints}\right)\\
&-2 \times \left( \text{Quadratic constraint} \right)\\
& - \left( \text{number of parameters of outer-Aut} \right)\\
& -\left( n \right)
\end{split}\nonumber
\end{equation}
where $n\equiv$ dim(inner-Aut) - number of independent linear
constraints.

In our case the number of independent linear constraints is 3, and
the dimension of the inner-Aut is 2, so $n\,=\,-1$. The constants
that appear at the outer-Aut are 2 and obviously the number of
$\gamma_{\alpha \beta}$ is 6. Thus, the expected maximal number of
essential constants is indeed \textbf{3}, by both ways of counting .

Despite the relatively simple form of (\ref{final u III}), its
general solution is, to the best of our knowledge, not known.
However, we have managed to obtain a partial solution in the
parametric form \be\label{para final u III}
u(\xi)=\frac{4\,(1+2\,e^{2\,\xi})^{3/2}}{3\, (1+e^{2\,\xi})^2} &
\rho=\frac{1}{3}\,(5+sech^2\xi) \nonumber \ee which makes the
functions  $u_1,\,u_2,\,u_4$ read as \be
u_1(\xi) & = & k_1+\xi+\ln{cosh\xi}\nonumber\\
u_2(\xi) & = & k_2+\ln{sech\xi}-\frac{1}{2}\,\ln{(cosh2\xi+2)}\nonumber \\
u_4(\xi) & = &
k_4-\ln{cosh\xi}+\frac{1}{2}\,\ln{(cosh2\xi+2)}\nonumber\ee and
the lapse
$(N)^2=\frac{e^{k_1+2\,\xi}\,(e^{2\,\xi}+1)}{4\,(2\,e^{2\,\xi}+1)}
$. The ensuing metric, after the usual purification with the
constant automorphisms and a shift in $\xi$, is given by
:\be\label{metric u4 not con} ds^2 & = &\kappa^2\,\left(
-\frac{e^{2\,\xi}\,(e^{2\,\xi}+1)}{4\,(2\,e^{2\,\xi}+1)}\,d\xi^2+
\frac{e^\xi}{4}\,cosh\xi\,dx^2+e^{-2x+\xi}(cosh2\xi+2)\,sech\xi\,dy^2
\right.
\nonumber \\
\nonumber \\
& &
\left.\ph{-\frac{e^{2\,\xi}\,(e^{2\,\xi}+1)}{4\,(2\,e^{2\,\xi}+1)}\,}
+\,e^\xi\,sech\xi\,dz^2+2e^{-x+\xi}\,sech\xi\,dy\,dz \right) \ee

As we have already remarked, this metric  does not admit  a
Homothety and therefore the constant $\kappa$ is essential. It does
not satisfy the invariant relation (\ref{invar1}), and it is not a
pp-wave. Therefore we conclude that it is inequivalent to
(\ref{macdiag}) or (\ref{metric u4=con}). This mono-parametric
family belongs to the Kinnersley vacuum solutions \cite{Kinn}. It is
quite interesting that it also admits a fourth killing vector field
\be\label{newkil}
\xi_4=-16\,y\,\partial_x+(e^{2\,x}-8\,y^2)\,\partial_y-2\,e^x\,\partial_z
\ee which produces with (\ref{killing}) the following table of
(non-vanishing) commutators: \begin{equation}\label{commu2}
[\xi_1,\xi_3]=\xi_1,\,[\xi_1,\xi_4]=-16\,\xi_3,\,[\xi_3,\xi_4]=\xi_4
\end{equation}

The isotropy group inferred from the above algebra (see the last
commutator) is a $G_1$ spatial rotation.

 Curiously enough, this algebra is equivalent to (\ref{commu1})
as a simple scaling of $\xi_1,\,\xi_4$ by $2\sqrt{2}$ shows. Of
course, the multiply transitive character of the action of the
underlying group on the corresponding $V_3\,'s$ allows for these,
and thus for the space-times in which they are embedded, to be
inequivalent.

Again, the extra symmetry emerged in the course of the investigation
of the solution space. Of course it must have something to do with
the particular nature of the solution, but it was not set as a
starting point.

\subsection{Preview for other Bianchi Types}
The method described in the previous sections can be applied to
other Types as well. The general pattern is s similar to that of
Type III : The pp-wave solutions ( for Types admitting such
geometries) occupy one part of the solution space, the other known
solutions reside on another part, and the unknown part of the
solution space is always described by an ODE strikingly similar to
(\ref{final u III}), namely : \be\label{final eq}
\ddot{u}^2=\frac{(-1+\dot{u}^2)^2}
{(\kappa+\lambda\,\rho)\,(\rho^2-u^2-1)}\ee Details will be
included in a forthcoming work. As indicative examples we give the
form of the ODE for Types $IV$ and $VII_h$:

\textbf{Type IV}

 \be\label{final u IV}
\ddot{u}^2=\frac{(-1+\dot{u}^2)^2}
{(\kappa+\lambda\,\rho)\,(\rho^2-u^2-1)} & \kappa=-6,\,\lambda=6
\ee

\textbf{Type $ \mathbf{VII_h}$}

 \be\label{final u VΙ}
\ddot{u}^2=\frac{(-1+\dot{u}^2)^2}
{(\kappa+\lambda\,\rho)\,(\rho^2-u^2-1)}  &
\kappa=-6+\frac{4}{h^2},\,\lambda=-6\ee

and of course

\textbf{Type III}

 \be
\ddot{u}^2=\frac{(-1+\dot{u}^2)^2}
{(\kappa+\lambda\,\rho)\,(\rho^2-u^2-1)} & \kappa=-10,\,\lambda=6
\ee

\section{Discussion}
When one is trying to solve Einstein's Equations in cosmology, one
has to deal with a nonlinear system of coupled, \emph{ordinary
differential} equations. The strategy that is frequently used, is to
simplify the system by choosing,  a convincing form for the scale
factor matrix, usually obtained by an a priori assumption of extra
symmetry (e.g. $\gamma_{\alpha\beta}=diag{(a(t),b(t),c(t)}$) and
then try to solve it, hoping to find \emph{some} solution. Clearly,
this procedure can by no means  guarantee  access to the full space
of solutions for the problem at hand. In this work we have presented
a method for solving Einstein's Field Equations in the case of
vacuum Bianchi Geometries. The main idea is to consider the Group of
constant Automorphisms, which emerges as the residual freedom left
after the time dependent A .I.D.'s (\ref{gaugetrans}),
(\ref{intcon}) have been used to set the shift $N^\alpha $ to zero,
as a Lie point-symmetry of the EFE's. In a step-by-step procedure
one can bring some of  the generators of this group in normal form
and simplify the rest, thereby reducing the order of the system of
equations. Which of the generators, and how,  can be utilized in
each step  depends upon the characteristics of their Lie Algebra
(abelian, solvable etc.). It is also important that the information
gained at a particular level must be used and, in fact, may be vital
for the implementation of the next step. The method is, by
construction, sweeping out all possible solutions, since no ad-hoc
assumption has been made. Therefore, \emph{if successfully applied
to a given Bianchi Cosmology}, it will result in the cartography  of
the entire space of solutions.

The successful application of the procedure to Bianchi Type III
resulted in the recovery of all known solutions without prior
assumption of any extra symmetry ((\ref{macdiag}),(\ref{metric
u4=con})), the enclosure of the entire unknown part of the solution
space into a single, second order ODE in terms of one dependent
variable (\ref{final u III}), and a partial solution to this ODE. It
is of interest that the solution space is naturally partitioned into
three distinct disconnected pieces. Of great importance may be
considered the fact that a strikingly similar ODE describes the
unknown part of the solution space for other lower Bianchi Types.
For Types VIII, IX there remain no rigid automorphisms after the
shift has been set to zero and the constant rotations have been used
to diagonalize the scale-factor matrix. However there is the scaling
symmetry $Y_2$ that can serve as a starting point. This issue, along
with the presentation of the detailed cartography for the lower
Types is in our immediate scopes. Finally the method can be extended
towards either the inclusion of matter content, or in 4+1 Spatially
Homogeneous Cosmologies.

 \textbf{Acknowledgements}\\
The authors wish to thank Dr. G.O. Papadopoulops for enlightening
discussions concerning the invariants and Dr. P. Apostolopoulos for
bringing to their attention the existing solutions and discussions
on them. The authors are also indebted to prof. M.A.H. MacCallum
who, in a private communication pointed out that the solution
(\ref{metric u4 not con}) belongs to Kinnersley's family and is not
new. The project is co-funded by the European Social Fund and
National Resources - (EPEAEK II) PYTHAGORAS II .



\newpage
\begin{thebibliography}{99}
\bibitem{Schuk}
O. Heckman and E. Sch\"{u}cking, \textit{Relativistic Cosmology in
Gravitation (an introduction to current research)} edited by L.
Witten, Wiley
(1962)\\
\bibitem{Harvey}
A. Harvey, Jour. Math. Phys. \textbf{20}, 251 (1979)

\bibitem{ChrDim}
T. Christodoulakis, G. O. Papadopoulos and A. Dimakis, Jour. Phys.
A: Math. Gen. \textbf{36}, 427 (2003)

\bibitem{jantzen}
R. T. Jantzen Comm. Math. Phys. \textbf{64} (1979)  211; Jour. Math.
Phys. \textbf{23}, 1137 (1982); C. Uggla, R.T. Jantzen and K.
Rosquist, Phys. Rev D \textbf{51}(1995),5525\\
\bibitem{ashtekar}
J. Samuel and A. Ashtekar  , Class. Quan. Grav. \textbf{8}, 2191 (1991)\\
\bibitem{JMP} T. Christodoulakis, G. Kofinas, E. Korfiatis, G.O. Papadopoulos, A. Paschos \\
J.Math.Phys.  \textbf{42}, 3580-3608 (2001)

\bibitem{Henneaux} O. Coussaert and M. Henneaux, Class. Quantum Grav.
 \textbf{10}, 1607 (1993)

\bibitem{MacBook} \textit{ "Exact Solutions of Einstein's Field
Equations"} (Second Edition),H. Stephani, D. Kramer, M.MacCallum,C.
Hoenselaers and E. Hertl, Cambriodge  Monographs on Mathematical
Physics, CUP, Cambridge (2003)

\bibitem{Kinn}Kinnersley, W. J.Math.Phys.  \textbf{10}, 1195 (1969)

\bibitem{EM} G.F.R. Ellis and M.A.H. MacCallum,Commun.Math.Phys.  \textbf{12}, 108 (1969)


\bibitem{CMP} T. Christodoulakis, E. Korfiatis and G.O. Papadopoulos \\
Commun.Math.Phys.  \textbf{226}, 377-391 (2002)
\bibitem{Stephani} \textit{ "Differential Equations: Their Solution using Symmeteries"},
H. Stephani, Edited by M.A.H. MacCallum, Cambridge University
Press, Cambridge (1989)

\bibitem{Olver} See e.g. Peter J. Olver, \textit{``Applications of Lie Groups to Differential Equations''},\\
Springer, Graduate Texts in Mathematics 107, (2000)

\bibitem{Ryan} \textit{"Homogeneous Relativistic Cosmologies"}, M.P.
Ryan Jr. and L.C. Shepley, Prionceton University Press, Princeton
(1975)

\bibitem{Wainr} \textit{"Dynamical Systems in Cosmology"}, Edited
by J. Wainwright and G.F.R. Ellis, Cambridge University Press,
Cambridge (1997)

\bibitem{Siklos} S.T.C. Siklos, J. Phys. A: Math. Gen. \textbf{14}, 395-409 (1981)

\bibitem{Schmidt} H.J. Schmidt, Int. J. of Theor. Physics, \textbf{37(2)}, 691 (1998)

\bibitem{Kundt} W. Kundt, in "Recent Developments in General Relativity", p. 307, Pergamon Press, New York
(1962);"Les Theories Relativistes de la Gravitation", pp. 155,
Centre National de la Recherche Scientifique, Paris, (1962)

\bibitem{Geroch} R. Geroch, Annals of Physics, \textbf{48}, 526-540 (1968)

\bibitem{ChrHerv}
T. Christodoulakis, S. Hervik and G. O. Papadopoulos,  Jour. Phys.
A: Math. Gen. \textbf{37}, 4039 (2004)





\end{thebibliography}
\end{document}